\documentclass[fleqn,usenatbib]{mnras}
\usepackage{newtxtext,newtxmath}
\usepackage[T1]{fontenc}

\DeclareRobustCommand{\VAN}[3]{#2}
\let\VANthebibliography\thebibliography
\def\thebibliography{\DeclareRobustCommand{\VAN}[3]{##3}\VANthebibliography}
\usepackage{graphicx}
\usepackage{dcolumn}
\usepackage{comment}
\usepackage{bm}
\usepackage{hyperref}
\usepackage{float}
\usepackage{amsmath}
\usepackage{mathtools}
\usepackage{xcolor}
\usepackage{subcaption}

\usepackage[normalem]{ulem}

\newcommand{\Myr}{\,{\rm Myr}}
\newcommand{\Gyr}{\,{\rm Gyr}}

\newcommand{\kpc}{\,{\rm kpc}}
\newcommand{\kpcGyr}{\,{\rm kpc}\,{\rm Gyr}^{-1}}

\newcommand{\kms}{\,{\rm km}\,{\rm s}^{-1}}
\newcommand{\Msun}{\,{\rm M}_\odot}

\newcommand{\rmG}{{\rm G}}
\newcommand{\e}{\,\mathrm{e}}
\newcommand{\sech}{\,\mathrm{sech}}
\newcommand{\drm}{\mathrm{d}}
\newcommand{\pd}{{\partial}}
\newcommand{\rhoc}{\rho_{\rm c}}
\newcommand{\hPhi}{\hat{\Phi}}
\newcommand{\hf}{\hat{f}}

\newcommand{\mi}{\mathrm{i}}
\def\p{\partial}

\newcommand{\nn}{\nonumber}

\newcommand{\techo}{t_\mathrm{echo}}

\usepackage{accents}

\newcommand\CHrem{\bgroup\markoverwith{\textcolor{red}{\rule[0.5ex]{2pt}{0.9pt}}}\ULon}

\newcommand\RCrem{\bgroup\markoverwith{\textcolor{cyan}{\rule[0.5ex]{2pt}{0.9pt}}}\ULon}

\title[Galactic echoes]{Galactic echoes}

\author[R.~Chiba et al.]{
Rimpei Chiba$^{1,2}$\thanks{E-mail: rimpei-chiba@g.ecc.u-tokyo.ac.jp},
Jupiter Ding$^{3,4}$,
Chris Hamilton$^{5}$,
Matthew W.~Kunz$^{4,6}$ and
Scott Tremaine$^{5}$
\\
$^{1}$ Department of Astronomy, Graduate School of Science, The University of Tokyo, 7-3-1 Hongo, Bunkyo-ku, Tokyo, 113-0033, Japan \\
$^{2}$ Canadian Institute for Theoretical Astrophysics, University of Toronto, 60 St. George Street, Toronto, ON M5S 3H8, Canada \\
$^{3}$ CIERA and Department of Physics and Astronomy, Northwestern University, 2145 Sheridan Road, Evanston, IL 60208, USA \\
$^{4}$ Department of Astrophysical Sciences, Princeton University, 4 Ivy Lane, Princeton, NJ 08544, USA \\
$^{5}$ Institute for Advanced Study, Einstein Drive, Princeton, NJ 08540, USA \\
$^{6}$ Princeton Plasma Physics Laboratory, Princeton, NJ 08543, USA
}

\date{Accepted XXX. Received YYY; in original form ZZZ}
\pubyear{2025}

\begin{document}
\graphicspath{{./}{figures/}}
\label{firstpage}
\pagerange{\pageref{firstpage}--\pageref{lastpage}}
\maketitle

%%%%%%%%%%%%%%%%%%%%%%%%%%%%%%%%%%%%%%%%%%%%%%%%%%%%%%%%%%%%%%%%%%%%%%%%%%%%%%%%%%%%%%%%%%%%%%%%%%%%

\begin{abstract}
\textit{Gaia} has revealed a variety of substructures in the phase space of stars in the Solar neighborhood, including the vertical `Snail' in $(z,v_z)$ space. Such substructures are often interpreted as the incompletely phase-mixed response of the disc stars to a single perturbation, such as an impulsive encounter with a satellite galaxy. In this paper we consider the possibility that such structures contain manifestations of phase space \textit{echoes}. First established in plasma physics in the 1960s, echoes arise when a collisionless system is perturbed \textit{twice}: the macroscopic responses to both perturbations mix to small scales in phase space, whereupon they couple nonlinearly, producing a third macroscopic `echo' response without the need for a third perturbation. We derive the galactic analogue of the plasma echo theory using angle-action variables and apply it to a one-dimensional model of vertical motion in the Milky Way. We verify the predicted echo behavior using idealized test particle simulations, both with and without the inclusion of diffusion through orbital scattering off molecular clouds. While we conclude that the \textit{Gaia} Snail itself is unlikely a (pure) echo effect, the basic physics we uncover is sufficiently generic that we expect phase-space echoes to be common in disc galaxies.
\end{abstract}


%%%%%%%%%%%%%%%%%%%%%%%%%%%%%%%%%%%%%%%%%%%%%%%%%%%%%%%%%%%%%%%%%%%%%%%%%%%%%%%%%%%%%%%%%%%%%%%%%%%%

%%%%%%%%%%%%%%%%%%%%%%%%%%%%%%%%%%%%%%%%%%%%%%%%%%%%%%%%%%%%%%%%%%%%%%%%%%%%%%%%%%%%%%%%%%%%%%%%%%%%

\section{Introduction}
\label{sec:Introduction}

The \textit{Gaia} `Snail' \citep{antoja_dynamically_2018} is a one-armed spiral structure in the density of stars in the Solar neighborhood when viewed in vertical phase-space $(z,v_z)$. Here, $z$ is the vertical position of a star normal to the galactic mid-plane and $v_z$ is its vertical velocity. The existence of the Snail demonstrates that the Milky Way is not in phase-mixed equilibrium. In fact, Snail-like structures manifest not only in stellar density, but also in stellar labels like age and metallicity \citep{Frankel2024Iron}. Snails almost certainly result from the incomplete phase-mixing of the stars' distribution function (DF) following dynamical perturbations of some kind, but what precisely were/are these dynamical perturbations remains an unanswered question (see the review by \citealt{hunt2025milky} and references therein).

The simplest answer is that the Snail originated from a collision between the Milky Way and the Sagittarius dwarf galaxy roughly 300 million to 900 million years ago \citep[e.g.,][]{antoja_dynamically_2018,binney_origin_2018,Laporte2019Footprints,bland2021galactic,Asano2025Ripples}. However, several numerical studies have suggested that the data are not reproducible by the impact of Sagittarius alone \citep[e.g.,][]{bennett2022exploring}. Moreover, the Snail exhibits significant variation across different radii and azimuths, which is difficult to account for with a single perturbation \citep{hunt_multiple_2022,frankel_vertical_2023,Antoja2023GaiaDR3}. Motivated by these findings, \cite{tremaine_origin_2023} proposed an alternative scenario in which the Snail arises from the cumulative effect of many stochastic encounters with dark matter subhaloes as they pass through the disc. \cite{Gilman2025Dark} showed that such multiple encounters can indeed produce strong phase spirals, though only with a subhalo abundance several times higher than that predicted by cosmological simulations.

Here, we explore yet another type of satellite-disc encounter mechanism, which sits in-between the single-encounter and the many-encounter scenarios mentioned above. We consider the response of the disc to just \textit{two} successive, short-lived perturbations. We examine the subsequent nonlinear coupling between the responses to these two perturbations, and ask in particular whether the observed Snail can arise as a galactic analogue of the famous \textit{plasma echo}. This is the phenomenon by which a twice-perturbed plasma `spontaneously' exhibits a third macroscopic response because of a delayed nonlinear coupling between the first and second responses \citep{gould_plasma_1967}. 

More precisely, the simplest echoes arise when one applies two successive, impulsive kicks to a homogeneous box of collisionless plasma. Soon after each kick, the perturbed phase-space DF gives rise to a macroscopic density response. Then, the perturbed DF phase mixes to ever smaller scales in velocity space, in the same way that the Snail is expected to mix to ever smaller, and eventually unobservable, scales in frequency space \citep{tremaine_origin_2023}. Since macroscopic quantities like density are calculated via integrals over velocity, this mixing results in density perturbations that decay with time. However, since the plasma is collisionless, the information contained in the phase-mixed DF cannot be lost. In particular, the two wound-up responses can nonlinearly couple at small scales in phase space and `unwind' themselves, such that at some later time a piece of the perturbed DF ceases briefly to be rapidly oscillating in velocity space. This results in a third density perturbation---the so-called `echo'---without the need for a third kick. The echo phenomenon was predicted theoretically over half a century ago \citep{gould_plasma_1967} and was almost immediately confirmed experimentally \citep{malmberg_plasma_1968}.

Galaxies are also (nearly) collisionless systems in which external `kicks' (from dwarf galaxies, dark matter subhaloes, and the like) can produce macroscopic phase-space responses which then phase mix away, and the Snail is usually thought to be one such response. The question naturally arises as to whether a plasma echo-like effect can exist, and whether it is observable, in galaxies like the Milky Way. The purpose of this paper is to investigate this analogy. The main technical difficulty is that unlike the unperturbed trajectories in a homogeneous plasma, the unperturbed orbits of stars are not straight lines, so the natural variables to use are angle-action coordinates $(\boldsymbol{\theta}, \boldsymbol{J})$ rather than position and velocity $(\boldsymbol{x}, \boldsymbol{v})$.  In particular, the potential perturbation $\delta \phi(\bm{x})$ in these variables is then a function of both the canonical coordinate and its conjugate momentum $\delta \phi(\boldsymbol{\theta}, \boldsymbol{J})$. While this complicates the formalism, we will see that it does not fundamentally change the echo phenomenon and that we can indeed have galactic echoes that resemble the Snail.

The plan for the rest of this paper is as follows. In \S\ref{sec:Derivation} we rederive the echo phenomenon using angle-action variables, motivated by a simple one-dimensional model of vertical motion in the Milky Way. In \S\ref{sec:test_particle_simulation} we verify the echo theory using idealized test-particle simulations, both with and without the effects of molecular cloud scattering. In \S\ref{sec:Discussion} we discuss the implications of galactic echoes for the observed \textit{Gaia} Snail and for galactic dynamics more generally. We summarize in \S\ref{sec:Summary}.

%%%%%%%%%%%%%%%%%%%%%%%%%%%%%%%%%%%%%%%%%%%%%%%%%%%%%%%%%%%%%%%%%%%%%%%%%%%%%%%%%%%%%%%%%%%%%%%%%%%%

\section{Theory}
\label{sec:Derivation}

We introduce the basic framework behind the echo theory and establish our notation in \S\ref{sec:Basic_Framework}. Then in \S\ref{sec:echo} we derive the galactic echo effect in the simplest case, in which we excite the disc twice impulsively, and calculate to second order in perturbation theory, ignoring both self-gravity and `collisions' (small-scale diffusion due to, e.g., molecular cloud scattering). We discuss how these `nonideal' effects would modify the theory in \S\ref{sec:Key_Features}.

\subsection{Framework}
\label{sec:Basic_Framework}

We consider a stellar system described by a phase-space DF $f$ and gravitational potential $\Phi$. For simplicity, we consider motion in only one spatial dimension, $z$, and hence one velocity dimension, $v$. We assume that absent any perturbations, the gravitational field of our system is a known function $\Phi_0(z)$. Then the unperturbed Hamiltonian
\begin{equation}
    H_0(z, v) = \frac{1}{2} v^2 + \Phi_0(z)
    \label{eq:unperturbed_Hamiltonian}
\end{equation}
 is integrable, meaning that we can define the angle-action variables
\begin{align}
  \theta \equiv \Omega(J) \int_0^z \frac{\drm z}{v}\quad{\rm and}\quad J \equiv \frac{1}{2\pi} \oint \drm z ~v,
  \label{eq:angle_action}
\end{align}
where $\Omega(J)$ is the orbital frequency
\begin{align}
  \Omega(J) \equiv 2 \pi \left( \oint \frac{\drm z}{v} \right)^{-1}.
  \label{eq:frequency}
\end{align}
By construction, the Hamiltonian $H_0$ is a function of the action alone, $H_0=H_0(J)$, and the frequency is $\Omega(J)=\partial H_0 / \partial J$. 

The DF evolves according to the collisionless Boltzmann equation (CBE):
\begin{align}
  \frac{\drm f}{\drm t} = \frac{\pd f}{\pd t} + \frac{\pd f}{\pd \theta} \frac{\pd H}{\pd J} - \frac{\pd f}{\pd J} \frac{\pd H}{\pd \theta} = 0,
  \label{eq:CBE}
\end{align}
where $H = v^2/2+\Phi = H_0 + (\Phi-\Phi_0)$ is the full Hamiltonian.

\subsection{Galactic echo from second-order perturbation theory}
\label{sec:echo}

To derive the galactic echo, we perturb our system and write the response as a power series in the perturbation parameter $\epsilon$:
\begin{align}
  H(\theta, J, t) &= H_0(J) + \epsilon \Phi_1(\theta, J, t) + \epsilon^2 \Phi_2(\theta, J, t) + \dots , \\
  f(\theta, J, t) &= f_0(J) + \epsilon f_1(\theta, J, t) + \epsilon^2 f_2(\theta, J, t) + \dots .
  \label{eq:f_expand}
\end{align}
Here, we ignore the self-gravity of the response, so that $\Phi_1$ consists only of the external perturbation, and higher-order perturbations in $\Phi$ are all zero, i.e., $\Phi_2=\Phi_3=\dots=0$.

Expanding the CBE \eqref{eq:CBE} as a power series in $\epsilon$, and collecting terms with equal powers of $\epsilon$, we find at the first and second orders that
\begin{align}
  &\epsilon^1:~\frac{\pd f_1}{\pd t} + \frac{\pd f_1}{\pd \theta} \frac{\pd H_0}{\pd J} - \frac{\pd f_0}{\pd J} \frac{\pd \Phi_1}{\pd \theta} = 0,
    \label{eq:CBE_expand_1}
    \\
  &\epsilon^2:~\frac{\pd f_2}{\pd t} + \frac{\pd f_2}{\pd \theta} \frac{\pd H_0}{\pd J} + \frac{\pd f_1}{\pd \theta} \frac{\pd \Phi_1}{\pd J} - \frac{\pd f_1}{\pd J} \frac{\pd \Phi_1}{\pd \theta} = 0.
  \label{eq:CBE_expand_2}
\end{align}
Taking advantage of the $2\pi$ periodicity of the angle $\theta$, we expand the perturbations into Fourier series:
\begin{align}
  \Phi_j(\theta, J, t) &= \sum_{n} \hPhi_{j,n}(J, t) \e^{\mi n \theta}, 
  \\
  f_j(\theta, J, t) &= \sum_{n} \hf_{j,n}(J, t) \e^{\mi n \theta},
  \label{eq:Fourier_expansion_theta}
\end{align}
where $n$ is an integer, and $j=1,2,\dots$ denotes the order of the perturbation. 
Inserting \eqref{eq:Fourier_expansion_theta} into equations \eqref{eq:CBE_expand_1}--\eqref{eq:CBE_expand_2}  and Fourier transforming leads to
\begin{align}
  \frac{\pd \hf_{1,n}(J,t)}{\pd t} &+ \mi n \Omega \hf_{1,n}(J,t) = \mi n \frac{\pd f_0}{\pd J} \hPhi_{1,n}(J,t),
  \label{eq:CBE_f1n}\\
  \frac{\pd \hf_{2,n}(J,t)}{\pd t} &+ \mi n \Omega \hf_{2,n}(J,t) = \nonumber \\
  \mbox{} &- \mi \sum_{n'}\bigg[ n' \hf_{1,n'}(J,t) \frac{\pd \hPhi_{1,n-n'}(J,t)}{\pd J} \nn
  \\
  \mbox{} &- (n-n') \frac{\pd \hf_{1,n'}(J,t)}{\pd J} \hPhi_{1,n-n'}(J,t) \bigg].
  \label{eq:CBE_f2n}
\end{align}

\subsubsection{General solution}

Assuming the system is unperturbed at $t=0$, the general solution to the first-order equation (\ref{eq:CBE_f1n}) is 
\begin{align}
  \hf_{1,n}(J, t) = \int_0^t \drm t' \e^{- \mi n \Omega (t - t')} \mi n \frac{\pd f_0}{\pd J} \hPhi_{1,n}(J, t').
  \label{eq:fn_linear}
\end{align}
The general solution to the second-order equation (\ref{eq:CBE_f2n}) can then be obtained by substituting (\ref{eq:fn_linear}) and integrating:
\begin{align}
  \hf_{2,n}(J, t) 
  &= - \int_0^t \drm t' \e^{- \mi n \Omega (t - t')} \nonumber \\
  \mbox{} &\times \mi\sum_{n'} \bigg[n' \frac{\p \hPhi_{1,n-n'}(J,t')}{\p J}  - (n-n') \hPhi_{1,n-n'}(J,t') \frac{\p}{\p J} \bigg] \nonumber \\
  \mbox{} &\times \int_0^{t'} \drm t'' \e^{- \mi n' \Omega (t' - t'')} \mi n' \frac{\pd f_0}{\pd J}\hPhi_{1,n'}(J, t'').
  \label{eq:fn_2nd_0}
\end{align}
In principle our work is now done: for a given potential perturbation $\Phi_1$, the full solution for the DF to second order in perturbation theory is found as the sum of equations \eqref{eq:fn_linear} and \eqref{eq:fn_2nd_0}. However, at this stage the right hand side of \eqref{eq:fn_2nd_0} is rather opaque and does not make manifest the echo-like behavior of the solution that will arise when we apply two impulsive kicks. Since the echo is due to a temporary synchronization of phases, we now rewrite the solution \eqref{eq:fn_2nd_0} in a way that makes its phase dependence more clear. 
For notational ease, we first introduce the following operators:
\begin{equation}
  \p_J \equiv \frac{\p}{\p J}, \qquad
  D_J(t') \equiv \p_J + \mi n \frac{\p \Omega}{\p J}  (t - t').
  \label{eq:tilde_pdJ}
\end{equation}
We can then rewrite equation (\ref{eq:fn_2nd_0}) as  
\begin{align}
  \hf_{2,n}(J, t)
  &= \sum_{n'} n' \int_0^t \drm t' \int_0^{t'} \drm t'' \nonumber \\
  \mbox{} &\times \bigg[n' \pd_J \hPhi_{1,n-n'}(J,t') - (n-n') \hPhi_{1,n-n'}(J,t') D_J(t') \bigg] \nonumber \\
  \mbox{} &\times \frac{\pd f_0}{\pd J} \hPhi_{1,n'}(J, t'') \e^{- \mi \Omega [n t - (n-n') t' - n' t'']},
  \label{eq:fn_2nd}
\end{align}
where the operator $D_J$ acts on the entire third line. This form of the solution makes clear the crucial dependence of the second-order perturbed DF on the phase $\xi \equiv \Omega(J) [n t - (n-n') t' - n' t'']$. At large $t$ we typically expect $\e^{-\mi\xi}$ to be a very rapidly oscillating function of $J$. Then phase-space perturbations will be tightly wound, and corresponding macroscopic quantities like real-space density, whose calculation involves integrals over action $J$, will tend to be suppressed. The exception to this rule occurs around special times $t$ when $\xi$ vanishes, as we will see.

\subsubsection{Two impulsive kicks}

We now choose $\Phi_1$ to consist of two impulsive potential perturbations, at times $t=t_1$ and $t=t_2 > t_1$, respectively:
\begin{equation}
  \Phi_1(\theta, J, t) = \psi_1(J) \cos(n_1 \theta) \delta(t-t_1) + (1 \leftrightarrow 2),
  \label{eq:perturbation_two_impulse}
\end{equation}
where $(1 \leftrightarrow 2)$ denotes terms the same as the previous ones but with the indices switched between 1 and 2. Note that we have chosen each perturbation to be purely cosinusoidal in the angle variable $\theta$ with fixed wave numbers $n_1$ and $n_2$. A more general kick would be developed as a Fourier series in $\theta$ with both sine and cosine terms, but we opt for \eqref{eq:perturbation_two_impulse} to keep the exposition relatively simple.

The Fourier coefficients of the perturbed potential are 
\begin{equation}
  \hPhi_{1,n}(J, t) = \sum_{s=\pm} \psi_1(J) \frac{\delta_{n}^{s n_1}}{2} \delta(t-t_1) + (1 \leftrightarrow 2).
  \label{eq:Phin_two_impulse}
\end{equation}
Plugging (\ref{eq:Phin_two_impulse}) into (\ref{eq:fn_2nd}), we obtain 
\begin{align}
  &\hf_{2,n}(J, t)
   = \frac{1}{4} \sum_{n'} \sum_{s_a=\pm} \sum_{s_b=\pm} n' \int_0^t \drm t' \int_0^{t'} \drm t'' 
   \label{eq:fn_two_impulse} \\
  & \times \bigg\{\bigg[n' \pd_J \psi_1 - (n-n') \psi_1 D_J(t') \bigg] \delta_{n-n'}^{s_a n_1} \delta(t'-t_1) + (1 \leftrightarrow 2) \bigg\} \nn \\
  & \times \frac{\pd f_0}{\pd J} \bigg[\psi_1 \delta_{n'}^{s_b n_1} \delta(t''-t_1) + (1 \leftrightarrow 2) \bigg] \e^{- \mi \Omega [n t - (n-n') t' - n' t'']}. \nn
\end{align}
When expanded in full, the solution (\ref{eq:fn_two_impulse}) consists of many different terms, but they can be grouped into two classes, which we call an `uncoupled' class and a `coupled' class. Thus we write
\begin{equation}
  \hf_{2,n}(J, t)
  =  \hf_{2,n}^{\,\rm u}(J, t) +  \hf_{2,n}^{\,\rm c}(J, t),
\end{equation}
with the superscripts `u' and `c' signifying `uncoupled' and `coupled', respectively.

The `uncoupled' terms consist of all the terms on the right-hand side of \eqref{eq:fn_two_impulse} in which either $\psi_1$ appears twice or $\psi_2$ appears twice, without an appearance of both $\psi_1$ and $\psi_2$. A few lines of algebra show that the uncoupled terms can be written as 
\begin{align}
  \hf_{2,n}^{\,\rm u}(J, t) 
  &= \frac{1}{4} \sum_{s=\pm} \mathcal{H}(t - t_1) n_1^2 \bigg[ \delta_n^0 \bigg( \psi_1^2\frac{\pd^2 f_0}{\pd J^2}  + 2 \psi_1 \pd_J \psi_1 \frac{\pd f_0}{\pd J} \bigg) \nn \\
  \mbox{} &- \delta_n^{s 2n_1} \psi_1^2 \frac{\pd^2 f_0}{\pd J^2} \e^{- \mi n \Omega (t - t_1)}\bigg] + (1 \leftrightarrow 2),
  \label{eq:fn_harmonic}
\end{align}
where $\mathcal{H}$ denotes the Heaviside step function. These uncoupled terms are simply the nonlinear response of the original unperturbed DF to each individual kick. They contain no coupling between the two separate responses, and as such they do not give rise to any echo-like phenomena. 

The `coupled' terms are all the remaining terms in \eqref{eq:fn_two_impulse}, i.e., those in which both $\psi_1$ and $\psi_2$ appear. Because $t' > t''$ and $t_2 > t_1$, the only coupled terms that survive are
\begin{align}
  &\hf_{2,n}^{\,\rm c}(J, t)
  = \frac{1}{4} \sum_{s=\pm} \mathcal{H}(t - t_2) n_1
  \label{eq:fn_echo} \\ 
  &\hspace{1.8mm}\bigg\{\delta_{n}^{s(n_2+n_1)} \!\bigg[n_1 \pd_J \psi_2 - n_2 \psi_2 D_J(t_2) \bigg]\! \frac{\pd f_0}{\pd J} \psi_1 \!\e^{- \mi \Omega [n t - s (n_2 t_2 + n_1 t_1)]} \nn \\
  &+\delta_{n}^{s(n_2-n_1)} \!\bigg[n_1 \pd_J \psi_2 + n_2 \psi_2 D_J(t_2) \bigg]\! \frac{\pd f_0}{\pd J} \psi_1 \!\e^{- \mi \Omega [n t - s (n_2 t_2 - n_1 t_1)]} \bigg\}. \nn
\end{align}
We see that there are special times $t$ such that the complex exponential ceases to be a rapidly-varying function of $J$ through $\Omega(J)$. The first line in the curly bracket predicts that this occurs when $t=t_\mathrm{echo}^*$, where
\begin{equation}
 \techo^\ast \equiv \frac{n_2 t_2 + n_1 t_1}{n_2 + n_1} = t_2 - \frac{n_1 (t_2 - t_1)}{n_2 + n_1},
  \label{eq:echo_time_unphysical}
\end{equation}
while the second term predicts it at $t=t_\mathrm{echo}$, where
\begin{equation}
 \techo \equiv \frac{n_2 t_2 - n_1 t_1}{n_2 - n_1} = t_2 + \frac{n_1 (t_2 - t_1)}{n_2 - n_1}.
  \label{eq:echo_time}
\end{equation}
Let us assume without loss of generality that $n_1$ and $n_2$ are both positive. Then the echo time $\techo^\ast$ implied by \eqref{eq:echo_time_unphysical} occurs before $t_2$. Since $\hf_{2,n}^{\,\rm c}$ is only nonzero for $t>t_2$, there is no echo associated with this term. On the other hand, the echo time $\techo$ implied by \eqref{eq:echo_time} is greater than $t_2$ provided that $n_2 > n_1$. We may rewrite the associated `echo' term---the second term in the curly bracket of (\ref{eq:fn_echo})---by carrying out the derivative in $D_J$:
\begin{align}
  &\hf_{2,n}^{\,\rm echo}(J, t) 
  = \frac{1}{4} \sum_{s=\pm}\mathcal{H}(t - t_2) n_1 \delta_{n}^{s (n_2-n_1)}\nonumber \\
  \mbox{} &\times \bigg\{
  n_1 \psi_1 \pd_J \psi_2 \frac{\pd f_0}{\pd J} 
  + n_2 \psi_1 \psi_2 \frac{\pd^2 f_0}{\pd J^2} 
  + n_2 \psi_2 \pd_J \psi_1 \frac{\pd f_0}{\pd J} \nn \\
  \mbox{} &+ n_2 \psi_2 \psi_1 \frac{\pd f_0}{\pd J} [- \mi n \pd_J \Omega (t-t_{\rm echo}) + \mi n \pd_J \Omega (t - t_2)]
  \bigg\} \e^{- \mi n \Omega (t - t_{\rm echo})} \nn \\
  &= \frac{1}{4} \sum_{s=\pm} \mathcal{H}(t - t_2) n_1 \delta_{n}^{s (n_2-n_1)} \nn \\
  &\times \bigg[ \big(n_1 \psi_1 \pd_J \psi_2 + n_2 \psi_2 \pd_J \psi_1 \big) \frac{\pd f_0}{\pd J} + n_2 \psi_1 \psi_2 \frac{\pd^2 f_0}{\pd J^2} \nn \\
  &\hspace{5mm} + s \mi n_1 n_2 \psi_1 \psi_2 \pd_J \Omega (t_2 - t_1) \frac{\pd f_0}{\pd J} \bigg] \e^{- \mi n \Omega (t - t_{\rm echo})}.
  \label{eq:fn_echo_explicit}
\end{align}
Finally, substituting this expression into the Fourier representation of $f_2(\theta,J,t)$ (equation~\ref{eq:Fourier_expansion_theta}), we obtain our explicit echoing part of the DF in angle-action space:
\begin{align}
  &f_{2}^{\,\rm echo}(\theta, J, t)
  = \frac{1}{2} \mathcal{H}(t - t_2) \bigg\{ \bigg[ n_1 \big(n_1 \psi_1 \pd_J \psi_2 + n_2 \psi_2 \pd_J \psi_1 \big) \frac{\pd f_0}{\pd J} \nn \\
  &\hspace{13.3mm}+ n_1 n_2 \psi_1 \psi_2 \frac{\pd^2 f_0}{\pd J^2} \bigg] \cos \big[(n_2 - n_1) [\theta-\Omega (t - \techo)]\big] \nn \\
  & - n_1^2 n_2 \psi_1 \psi_2 \pd_J \Omega (t_2 - t_1) \frac{\pd f_0}{\pd J} \sin \big[(n_2 - n_1) [\theta - \Omega (t - \techo)] \big]\bigg\}.
  \label{eq:f_echo}
\end{align}
Again, a generic pair of perturbations would consist of a sum over terms $n_1$ and $n_2$ with arbitrary phases rather than the simplified form \eqref{eq:perturbation_two_impulse}, but the generalization of the above results is straightforward.

Equation (\ref{eq:f_echo}) shows explicitly that around the time $t=\techo$, the second-order perturbation to the DF consists of a piece that is \textit{not} rapidly oscillating in action space. This leads to a transient fluctuation---an `echo'---in macroscopic integrated quantities like real-space density, long after the original (first-order) perturbations have phase mixed away. In the context of the Milky Way, this suggests that loosely wound `Snails' can appear in phase space some time after the original Snails have become tightly wound. We demonstrate this numerically in \S\ref{sec:test_particle_simulation}.

\subsubsection{Discussion}
\label{sec:Interim_Discussion}

The key conclusion from our calculation is that a system perturbed by two successive, impulsive kicks with wave numbers $n_1$ and $n_2$ will exhibit a second-order `echo' response at wave number $n=n_2-n_1$ if $n_2 > n_1$. The latter requirement encodes the physical fact that perturbations with larger $n$ phase mix more quickly, and so for the second perturbation to `catch up' with the first and nonlinearly couple to produce an echo, it must have a larger wave number.

Note that the second-order echo term (\ref{eq:f_echo}) involves two terms, the second of which is proportional to $t_2-t_1$. As a result, the \textit{longer} one waits in-between the two kicks, the \textit{stronger} the nonlinear echo will ultimately be! Physically, this is due to the fact that the second-order response $f_2$ depends on the gradient of the first-order response $f_1$ in action space (see equation \ref{eq:CBE_expand_2}), and the longer one waits for $f_1$ to phase mix, the sharper are its action-space gradients (the original Snails are `wound up' more).

Note also that the conclusions we have drawn so far are directly analogous to the plasma-kinetic case, in which $J$ would be replaced by momentum and $\theta$ by position. The major difference in the galactic case is that the potential perturbations are a function of both canonical coordinates \textit{and} momenta, so we have terms involving derivatives $\partial \psi/\partial J$ that would be zero in the plasma case. However, such non-zero derivatives only result in simple corrections to the amplitude of the echo (appearing on the first line in equation \ref{eq:f_echo}). They do not qualitatively alter the echo phenomenon, meaning that galactic and plasma echoes are closely analogous.

\subsection{Limitations and extensions}
\label{sec:Key_Features}

While the calculation we presented in \S\ref{sec:echo} captures the basic physics behind `galactic echoes', it is highly idealized in several respects. It is important to understand these idealizations because they will inform how we interpret the numerical simulations presented in \S\ref{sec:test_particle_simulation}.

The first major idealization we made is that the dynamics is \textit{perturbative}. There are often circumstances in which this perturbative assumption (and the corresponding truncation of perturbation theory at second order) breaks down even for relatively weak kicks (small $\psi_{i}$). This should not be surprising, since, as discussed in \S\ref{sec:Interim_Discussion}, the perturbative echo calculation gives a term proportional to $t_2-t_1$. Naively this proportionality suggests that it is possible for the second-order response to have the same (or even a larger) amplitude than the first-order response \eqref{eq:fn_linear}. Realistically, though, the perturbative expansion \eqref{eq:f_expand} would break down in such a scenario. On the other hand, one can perform a fully general, non-perturbative calculation of the echo effect in the special case that $\p_J\Omega$ is constant and both $\psi_1$ and $\psi_2$ are independent of $J$. We outline this calculation in Appendix~\ref{sec:nonlinear_saturation_echo}. 

The second major idealization is that we ignored the effect of \textit{diffusion}, namely the fact that small-scale inhomogeneities like molecular clouds will give stochastic perturbations to stars' orbits. Put in plasma language, we did not include any collision operator on the right hand side of \eqref{eq:CBE}. Following \citet{su_collisional_1968}, \citet{tremaine_origin_2023}, and \citet{chiba2025origin}, we expect that if we were to include sufficient diffusion, it would act to erase observable echo effects. Mathematically, one can show that each first-order response would  approximately decay $\propto \e^{-(t/t_{\rm d})^3}$, where $t_{\rm d}$ is a characteristic diffusion time.\footnote{This super-exponential decay arises from the diffusion term in the collision operator. When a friction term is included, the decay transitions to an exponential form beyond the collision time \citep{Banik2024Relaxation}.} Heuristically, we expect that if either $t_2-t_1$ or $t_\mathrm{echo}-t_1$ is longer than $t_{\rm d}$, then the echo will not manifest. We confirm this behavior with numerical simulations in \S\ref{sec:test_particle_simulation}.

Third, we have ignored the self-gravity of the perturbed stellar distribution. As shown by \cite{Darling2019Emergence} and \cite{widrow_swing_2023}, self-gravity may strengthen the amplitude of the Snail-like response to an impulsive perturbation, and extend the lifetime of a Snail well beyond that expected due to phase mixing alone. It is unclear what impact self-gravity would have on galactic echoes. On the one hand, we expect that if self-gravity leads to stronger first-order responses, it should lead to stronger second- and higher-order effects like echoes too. On the other hand, the $t_2-t_1$ dependence of the echo term arises precisely because nonlinear couplings feed off phase-space gradients, and under normal phase mixing these get sharper as $t_2-t_1$ increases. Since self-gravity will reduce the rate of phase mixing, it may make echoes weaker, not stronger. In a homogeneous plasma, one can perform a self-consistent calculation \citep{gould_plasma_1967,ONeil1968Temporal}, which predicts that collective effects amplify the echo while slightly modifying its growth and decay rates. However, to perform such a self-consistent calculation here would incur a significant mathematical penalty: not only would we need to Laplace transform from the time-domain into the complex frequency domain \citep{gould_plasma_1967}, but also, since we are dealing here with an inhomogeneous system, we would need to project our fluctuations onto a biorthogonal potential-density basis \citep[e.g.,][]{Kalnajs1976Biorthonormal,Weinberg1991Vertical,hamilton_kinetic_2024}. Alternatively, we could forego both of these transforms and just solve the linearized equations numerically \citep{widrow_swing_2023}. Since our aim here is to demonstrate the existence of an echo phenomenon rather than perform detailed modeling of the Solar neighborhood, we will not investigate the role of self-gravity further in this paper.

%%%%%%%%%%%%%%%%%%%%%%%%%%%%%%%%%%%%%%%%%%%%%%%%%%%%%%%%%%%%%%%%%%%%%%%%%%%%%%%%%%%%%%%%%%%%%%%%%%%%

\section{Test-particle simulations}
\label{sec:test_particle_simulation}

In this section, we test the theoretical ideas developed in the previous section using test-particle simulations of one-dimensional (vertical) motion in a galactic disc.

\subsection{Model}
\label{sec:model}

We model the unperturbed disc with a self-gravitating, isothermal slab described by the following potential-density pair:
\begin{align}
  \Phi_0(z) = 2 \sigma^2 \ln \left[ \cosh \left(\frac{z}{2h}\right)\right], \quad
  \rho_0(z) = \rhoc \sech^2 \left(\frac{z}{2h}\right),
  \label{eq:isothermal_slab_potential_density}
\end{align}
where $h$ is the scale height at large $z$, $\sigma$ is the velocity dispersion, and $\rhoc = \sigma^2 / (8\pi \rmG h^2)$ is the central density. By default, we set $h=0.2 \kpc$ and $\sigma = 20 \kpcGyr = 19.6 \kms$. 
For our unperturbed DF we take
\begin{align}
  f_0(z,v) = \frac{\rhoc}{(2 \pi \sigma^2)^{1/2}} \exp\left[-\frac{H_0(z,v)}{\sigma^2}\right],
  \label{eq:isothermal_slab_f0}
\end{align}
where $H_0$ is the unperturbed Hamiltonian (\ref{eq:unperturbed_Hamiltonian}).

We subject our disc to two impulsive perturbations caused by passing satellites or dark-matter subhalos. According to the linear response analysis by \cite{banik_comprehensive_2022,banik_comprehensive_2023}, such encounters can excite both dipole ($n=1$) and quadrupole ($n=2$) modes in the vertical phase-space distribution of stars. The relative amplitudes of these modes depend critically on the parameters of the disc-satellite encounter, in particular, the encounter speed, $v_{\rm p}$, and the angle at which the satellite crosses the disc, $\vartheta_{\rm p}$, where $\vartheta_{\rm p}=0$ corresponds to a perpendicular passage through the disc.

To illustrate how two perturbations with different wave numbers can generate an echo, we adopt a simple scenario in which the first satellite imparts a uniform kick $\Delta v_1(z) = \Delta V_1$ to stars at $t=t_1$, exciting primarily a dipole-like ($n_1=1$) perturbation. As a rough estimate, a satellite of mass $M_{\rm p} = 4 \times 10^9 \Msun$, velocity $v_{\rm p} = 300 \kpcGyr$, and impact parameter $b = 10 \kpc$ would impart a kick of order $\Delta V_1 \sim \rmG M_{\rm p}/bv_{\rm p} \sim 6 \kpcGyr$ (see \citealt{binney_origin_2018} for a more elaborate modelling). Later, at $t=t_2$, we assume that the disc encounters another satellite/subhalo which crosses near the Solar neighborhood at a grazing angle, thereby inducing a quadrupole-like ($n_2=2$) perturbation. We determine the $z$ dependence of the second kick by assuming that the perturber is a Plummer sphere, crossing the disc at $\vartheta_{\rm p} = \pi/2$. The net kick imparted to stars in the impulsive limit is then (see Appendix~\ref{sec:impulsive_satellite_perturbation})
\begin{align}
  \Delta v_2(z) = - \Delta V_2 \frac{2za}{z^2 + a^2}, \quad {\rm where} \quad \Delta V_2 = \frac{\rmG M_{\rm p}}{a v_{\rm p}},
  \label{eq:Dv2}
\end{align}
and $a$ is the core radius. The validity of the impulse approximation is examined in Appendix~\ref{sec:impulsive_satellite_perturbation}. In our fiducial setup, we adopt $t_1 = 0.2 \Gyr$, $t_2 = 0.8 \Gyr$, $\Delta V_1 = 6 \kpcGyr$, $\Delta V_2 = 6 \kpcGyr$ and $a=0.3 \kpc$. The corresponding mass of the second satellite/subhalo is $M_{\rm p} \sim 1.2 \times 10^8 \Msun$, although this estimate can be much larger if the satellite passes at a non-zero distance from the Solar neighborhood.

We also examine the echo formation in a more realistic noisy disc, in which stars receive many weak, stochastic perturbations from small-scale structures like molecular clouds.\footnote{See the recent $N$-body+hydrodynamic simulations by \cite{TepperGarcia2025Galactic} for the effect of clumpy, turbulent gas on the formation of phase spirals.} We model this small-scale diffusion as a Gaussian random walk of stars in $(z,v)$ space\footnote{Because scattering by molecular clouds is a local process, it may be more appropriate to kick stars only in velocity, rather than in both $z$ and $v$. However, this choice results in negligible differences (see Appendix~C of \citealt{chiba2025origin}).}: we apply independent random kicks $(\Delta z, \Delta v)$ to each star every $\Delta t = 10 \Myr$, where the kicks have zero mean, $\langle \Delta z \rangle = \langle \Delta v \rangle = 0$, and variance, $\langle (\Delta z)^2 \rangle = D_z \Delta t$, $\langle (\Delta v)^2 \rangle = D_v \Delta t$, where $D_z,D_v$ are the diffusion coefficients. We assume that the diffusion coefficients are constants and determine their values by assuming that the disc was initially razor thin (i.e., all stars have zero action initially) and was gradually heated over time $T$ to obtain its present mean height $\overline{z^2}$ and velocity dispersion $\overline{v^2}$. Thus, $D_z = \overline{z^2} / T \sim 3.62 h^2/T $ and $D_v = \overline{v^2} / T = \sigma^2 / T$. By default, we set $T=10 \Gyr$.

\subsection{Emergence of echo in the disc's vertical phase space}
\label{sec:eho_phase_space}

\begin{figure*}
    \centering
    \includegraphics[width=0.63\linewidth]{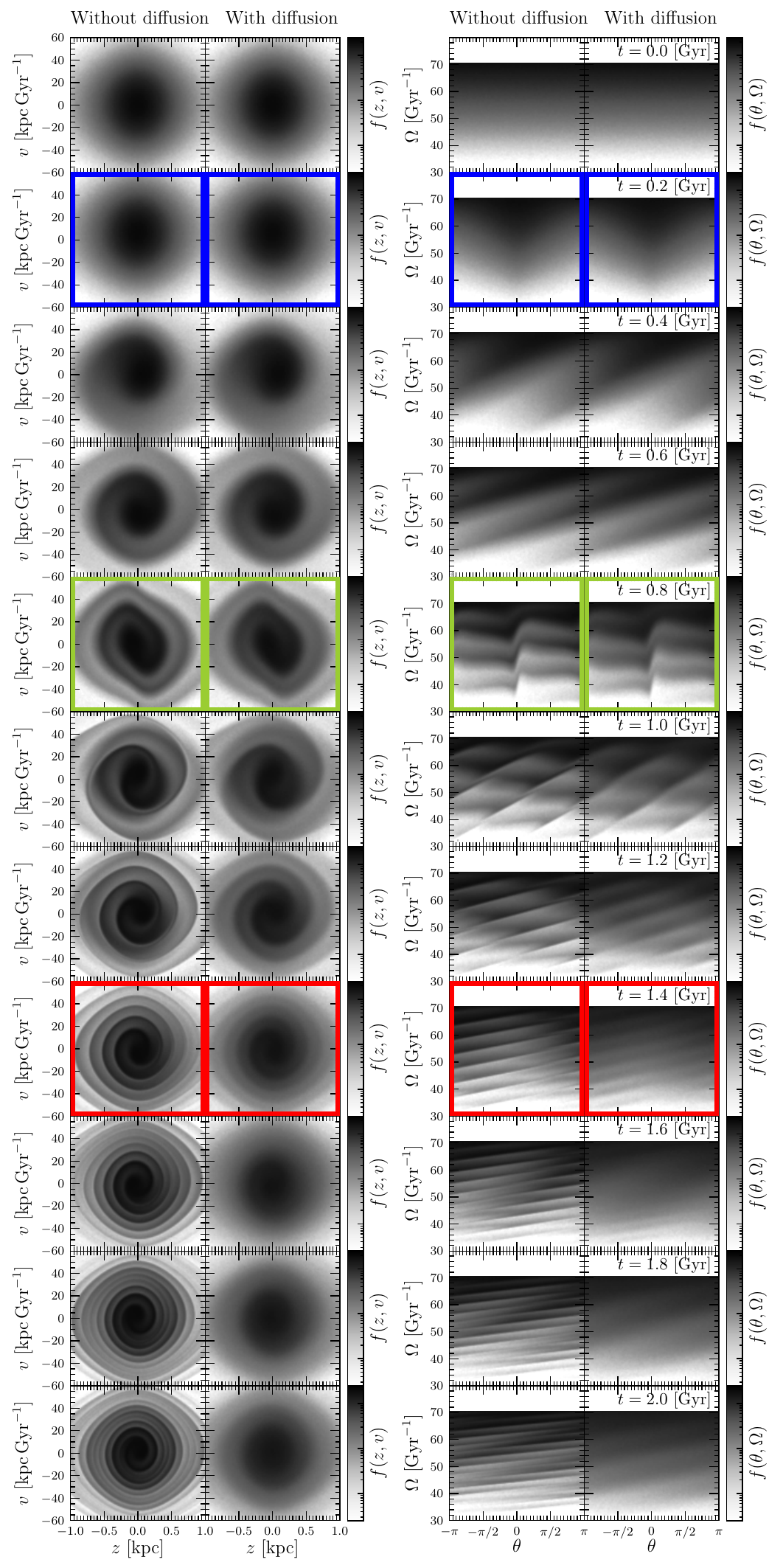}
    \caption{Phase-space evolution of an isothermal slab perturbed by two successive impulsive perturbations at $t_1=0.2 \Gyr$ (blue) and $t_2=0.8 \Gyr$ (green). Nonlinear coupling between the two perturbations results in a third phase spiral---the echo---which unwinds until $t=1.4 \Gyr$ (red) and then rewinds again.}
    \label{fig:echo_simulation_DF}
\end{figure*}

Fig.\,\ref{fig:echo_simulation_DF} shows the time evolution of the vertical phase-space distribution of the disc, which is hammered twice at $t_1=0.2 \Gyr$ (blue) and at $t_2=0.8 \Gyr$ (green), resulting in an echo at $t_{\rm echo} = (n_2 t_2 - n_1 t_1)/(n_2 - n_1) = 1.4 \Gyr$ (red). We plot the distribution in both the $(z,v)$ space (left two columns) and the $(\theta,\Omega)$ space (right two columns), and compare simulations with and without small-scale diffusion. In all simulations, we employ $10^8$ particles and compute the phase-space density on a $200\times 200$ regular mesh with a logarithmic scaling.

The first uniform kick (blue) shifts the entire distribution towards higher velocity, resulting in an overdensity at $\theta=0$ and underdensity at $\theta=\pm\pi$. This dipole ($n_1=1$) perturbation subsequently shears into a one-armed phase spiral in the $(z,v)$ plane. In the $(\theta,\Omega)$ plane, this spiral appears as an inclined, rectilinear stripe. The second antisymmetric kick (green) introduces a quadrupole ($n_2=2$) perturbation, which evolves into a two-armed phase spiral. In addition to generating its own response, the second kick also perturbs the pre-existing one-armed phase spiral. This second-order perturbation results in a new anti-wound spiral pattern, which appears as a linear pattern with negative inclination in the $(\theta,\Omega)$ plane. As time progresses, this pattern unwinds (phase `unmixes'), and at the predicted echo time (red), its phase fully aligns, giving rise to an echo. The echo subsequently phase mixes and forms a new one-armed phase spiral pattern.

\begin{figure}
    \centering
    \includegraphics[width=1\linewidth]{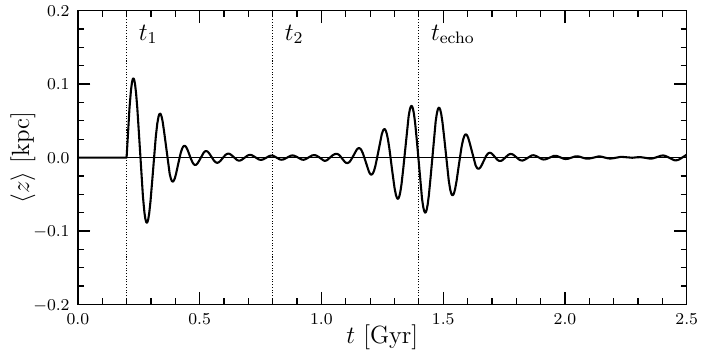}
    \includegraphics[width=1\linewidth]{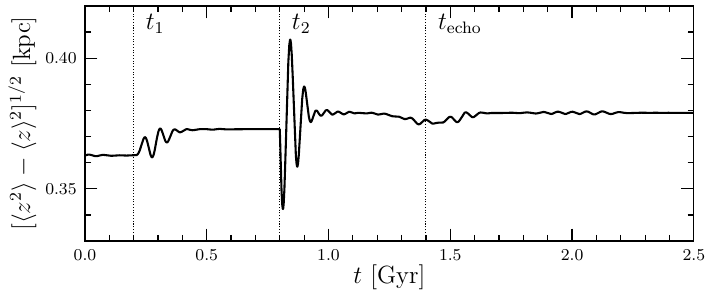}
    \includegraphics[width=1\linewidth]{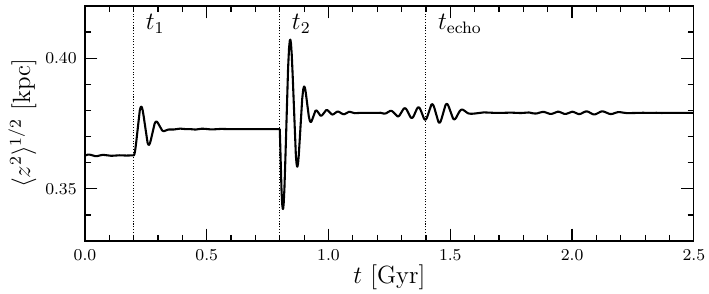}
    \caption{Time evolution of the mean height of the disc $\langle z \rangle$, its thickness 
    $[\langle z^2 \rangle-\langle z \rangle^2]^{1/2}$, and the rms vertical coordinate $\langle z^2 \rangle$,
    all in the absence of diffusion. The mean height undergoes damped oscillation after the first uniform kick at $t_1$ due to phase mixing. The second antisymmetric kick at $t_2$ does not directly affect the mean height, although it distorts the pre-existing one-armed phase spiral, generating a renewed oscillation (an echo) near $t_{\rm echo}$. The echo signals in the other panels are weaker (note the shrunken vertical axis).}
    \label{fig:meanz}
\end{figure}

The phase alignment at $t_{\rm echo}$ implies that the echo temporarily causes a fluctuation in macroscopic (integrated) quantities. To illustrate this, we plot in the top panel of Fig.\,\ref{fig:meanz} the time evolution of the mean height of the disc, $\langle z \rangle = \int {\rm d} z{\rm d} v f(z,v,t) z$. Following the first kick, the mean height undergoes a rapid oscillation, whose amplitude gradually decays due to phase mixing. The second kick induces a density perturbation symmetric about the disc mid-plane and therefore leaves the mean height unaffected. Later, as the echo unwinds and rewinds, the disc's mean height begins to oscillate again spontaneously, despite the lack of any external forcing. This demonstrates how the memory of earlier perturbations can be preserved in phase space and reappear later in macroscopic space. The remaining panels of Fig.\,\ref{fig:meanz} show a similar (though less dramatic) echo effect in the measurement of the disc's `thickness' $[\langle z^2 \rangle-\langle z \rangle^2]^{1/2}$ and in $\langle z^2 \rangle$ alone. The effect is less pronounced because the echo has most of its power at $n = \pm 1$, which does not contribute to even moments of $z$.

In the presence of small-scale diffusion (2nd and 4th columns of Fig.\,\ref{fig:echo_simulation_DF}), the phase spirals gradually decay over time. Because the effect of diffusion is proportional to the second derivative of the phase-space density \citep{tremaine_origin_2023,chiba2025origin}, older---and thus more tightly wound---phase spirals are erased more effectively. As a result, the first-order responses are largely smeared out at late times, while the second-order response (the echo) still persists (see also the bottom panel of Fig.\,\ref{fig:fit_model} below). This suggests that some observed phase-space features may in fact be (second-order) echoes rather than direct (first-order) responses to impulsive perturbations.

\subsection{Quantifying the echo}
\label{sec:echo_quantification}

To explore the possibility of observing an echo, we quantify the amplitudes and winding times of the phase spirals by fitting the following model to each snapshot \citep{frankel_vertical_2023}:
\begin{align}
  \frac{\delta f(\theta,J,t)}{\bar{f}(J,t)} = \sum_i A_i \cos n_i \left[\theta - \Omega(J) t_{{\rm w},i} - \theta_{0,i}\right],
  \label{eq:fit_model}
\end{align}
where $\delta f \equiv f - \bar{f}$ and $\bar{f}$ is the angle-averaged DF. The fitting parameters are the dimensionless amplitude $A$, the winding time $t_{\rm w}$, and the initial phase $\theta_0$. The sum runs over the three phase spirals ($i=1,2$, echo), yielding a total of 9 fitting parameters. The wave numbers are set to the dominant modes: $n_1=1$, $n_2=2$, and $n_{\rm echo}=n_2-n_1=1$. 

\begin{figure}
    \centering
    \includegraphics[width=1\linewidth]{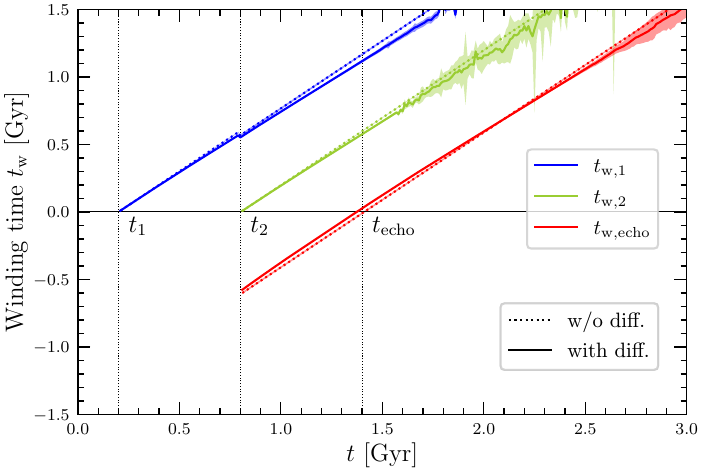}
    \includegraphics[width=1\linewidth]{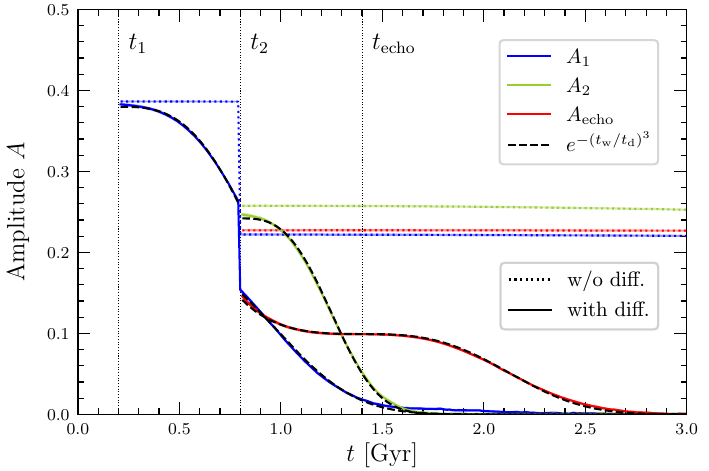}
    \caption{Winding time (top) and amplitude (bottom) of the phase spirals as a function of time. Blue and green mark those of the phase spirals directly induced by the first and second kick, respectively, while red represents the echo. Results both with (solid) and without diffusion (dotted) are plotted. The shaded regions around each line represent the 1-$\sigma$ uncertainty of the fit, which are invisibly small in the collisionless case (dotted). Since the amplitudes decay super-exponentially in the presence of diffusion (black dashed curves), the echo becomes much larger than the original phase spirals at late times.}
    \label{fig:fit_model}
\end{figure}

The upper panel of Fig.\,\ref{fig:fit_model} shows the time evolution of the phase spiral's winding time $t_{\rm w}$ with blue and green indicating those generated by the first and second kicks, respectively, and red representing the echo. As expected, the winding times increase almost linearly with time, regardless of the absence (dotted) or presence (solid) of small-scale diffusion. The winding times of the first-order responses are zero just after the kicks, whereas that of the echo starts negative and crosses zero at $t_{\rm echo}$, becoming positive thereafter, consistent with equation (\ref{eq:f_echo}). We observe a slight discontinuity of $t_{{\rm w},1}$ at $t_2$ due to the distortion of the one-armed phase spiral by the second kick. There is also a modest tendency for small-scale diffusion to decrease the inferred winding time: the winding rate slightly decreases when $t_{\rm w}>0$, and increases when $t_{\rm w} <0$.

The lower panel of Fig.\,\ref{fig:fit_model} plots the corresponding amplitudes of the phase spirals. In the absence of diffusion (dotted), the amplitudes are approximately constant, except at $t_2$ when the second kick distorts and weakens the pre-existing $n=1$ phase spiral (blue). In the presence of small-scale diffusion (solid), the amplitudes are reduced at a rate that depends on the winding time. As discussed briefly in \S \ref{sec:Key_Features}, when diffusion works in combination with phase mixing, perturbations in phase space decay super-exponentially with amplitude roughly $\propto$ $\exp[-(t_{\rm w}/t_{\rm d})^3]$, where $t_{\rm d}$ is the characteristic diffusion time \citep[e.g.,][]{su_collisional_1968,tremaine_origin_2023,chiba2025origin}. The black dashed curves in Fig.\,\ref{fig:fit_model} show that such a super-exponential fit agrees excellently with all simulations that include diffusion. The fitted values of $t_\mathrm{d}$ were $0.82,0.89,0.52,0.83\Gyr$ for $A_1(t<t_2),A_1(t>t_2),A_2,A_\mathrm{echo}$. The diffusion time of $A_2$ is shorter than others by a factor of about $0.6$, consistent with the scaling $t_\mathrm{d} \propto n^{-2/3}$ \citep[e.g.,][]{tremaine_origin_2023}, which yields $t_{\mathrm{d},n=2}/t_{\mathrm{d},n=1} = 2^{-2/3} \simeq 0.63$.

\begin{figure}
    \centering
    \includegraphics[width=1\linewidth]{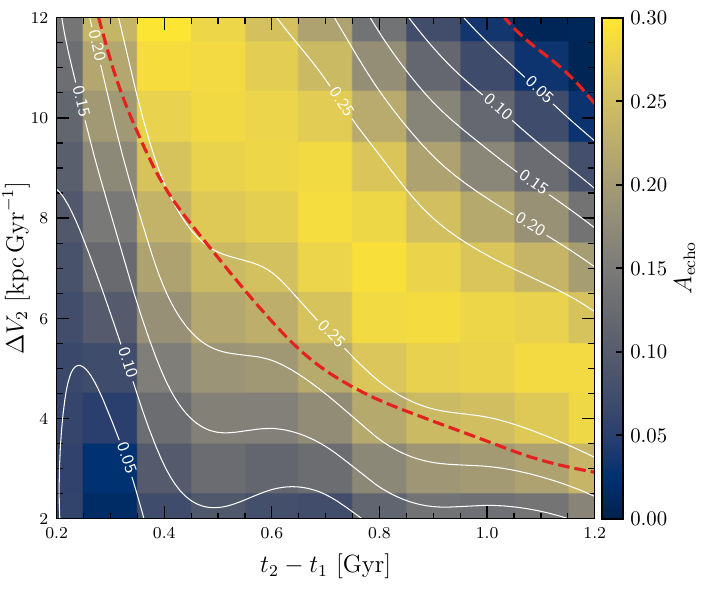}
    \caption{Amplitude of the echo, $A_{\rm echo}$, in the absence of small-scale diffusion, as a function of the time interval between the two large-scale impulsive kicks, $t_2-t_1$, and the strength of the second kick, $\Delta V_2$. The white curves are the contours of $A_{\rm echo}$ obtained via bivariate spline interpolation, and the red dashed curves mark the boundary within which $A_{\rm echo}$ exceeds $A_1$. The echo amplifies with both parameters but eventually saturates and begins to decline.}
    \label{fig:t2t1_dV2_Aecho_NoSSDiff}
\end{figure}

One of the key predictions of equation (\ref{eq:f_echo}) is that the amplitude of the echo increases with the time interval between the two impulses as well as their strength. We explore this is in Fig.\,\ref{fig:t2t1_dV2_Aecho_NoSSDiff}, which plots the amplitude of the echo, $A_{\rm echo}$, computed in the absence of small-scale diffusion, as a function of the time separation, $t_2-t_1$, and the strength of the second kick, $\Delta V_2$ (equation \ref{eq:Dv2}). We measured $A_{\rm echo}$ at the echo time, $t_{\rm echo}$, although the result is insensitive to this choice since $A_{\rm echo}$ remains constant in the absence of diffusion (see Fig.\,\ref{fig:fit_model}). We see that, for small values of $t_2-t_1$ and $\Delta V_2$, the echo amplitude increases with both parameters, consistent with equation (\ref{eq:f_echo}). However, as the parameters are made increasingly large, the echo amplitude eventually saturates and begins to decline, signaling the breakdown of second-order perturbation theory. The two red dashed curves mark the boundary within which the echo amplitude exceeds that of the original one-armed phase spiral. The saturation occurs approximately along a hyperbolic curve, consistent with the predicted scaling $\Delta V_2(t_2-t_1)$ from equation (\ref{eq:f_echo}). The underlying cause of this saturation is beyond the scope of the main text and is therefore deferred to Appendix~\ref{sec:nonlinear_saturation_echo}. As shown there, for even larger values of $t_2-t_1$ and $\Delta V_2$, the echo amplitude exhibits an oscillatory behavior. 

\begin{figure}
  \centering
  \includegraphics[width=1\linewidth]{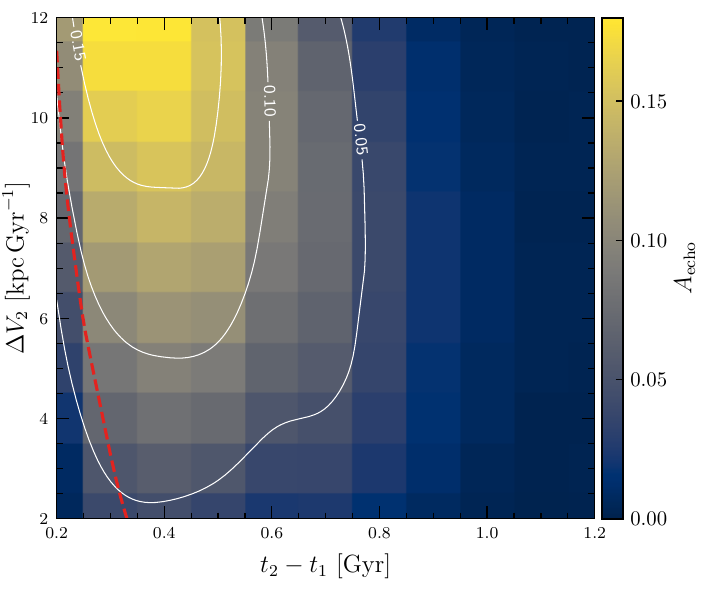}
  \caption{As in Fig.\,\ref{fig:t2t1_dV2_Aecho_NoSSDiff}, but with small-scale diffusion by, e.g., molecular cloud scattering.}
  \label{fig:t2t1_dV2_Aecho}
\end{figure}

In more realistic cases, where small-scale diffusion wipes out the echo signal over time, the echo amplitude may decline with increasing $t_2-t_1$ before reaching the non-perturbative regime. Fig.\,\ref{fig:t2t1_dV2_Aecho} plots the echo amplitude similar to Fig.\,\ref{fig:t2t1_dV2_Aecho_NoSSDiff}, but in the presence of small-scale diffusion. Here, we plot the amplitude of the echo when its winding time is $t_{\rm w} = 0.4 \Gyr$ as observed in the Milky Way \citep{frankel_vertical_2023,Antoja2023GaiaDR3}. The amplitude of the echo drops sharply with $t_2-t_1$ beyond $\sim 0.5 \Gyr$, implying that echoes in real noisy galaxies may not experience the saturation seen in the purely collisionless case. Our result suggests that, due to the diffusion-limited growth, the echo would be most pronounced when the two large-scale kicks are separated by roughly $0.4 \Gyr$. Note also that the red dashed curve, which marks the boundary beyond which $A_{\rm echo}$ exceeds $A_1$, has shifted significantly to the left compared to that in Fig.\,\ref{fig:t2t1_dV2_Aecho_NoSSDiff}. In other words, in the presence of small-scale diffusion it is \textit{easier} for the echo amplitude to dominate over the original responses.

\section{Discussion}
\label{sec:Discussion}

Theoretical studies of Milky Way dynamics often focus on explaining substructures in phase space like the \textit{Gaia} Snail \citep{antoja_dynamically_2018,hunt2025milky}. Most such studies employ either numerical simulations or linear (semi-)analytical theory \citep{binney_origin_2018,banik_comprehensive_2022,banik_comprehensive_2023,tremaine_origin_2023,widrow_swing_2023}, with the latter often serving as a foundation to interpret the former. However, \textit{nonlinear} mechanisms for producing these substructures have rarely been explored analytically, with the notable exception of in-plane orbital trapping \citep{monari2017distribution,binney2020trapped,chiba2021resonance}. 

Among the nonlinear phenomena that can give rise to distinct phase-space substructures, perhaps the simplest and most analytically tractable effect is a phase-space echo, which has been studied extensively in plasma physics but apparently never in the context of galactic dynamics. Our theoretical derivation (\S\ref{sec:Derivation}) and numerical simulations (\S\ref{sec:test_particle_simulation}) suggest that such echo effects may well exist in the vertical phase space of galactic discs. More generally, although we restricted ourselves here to a one-dimensional phase space and impulsive perturbations for simplicity, an extension of \S\ref{sec:Derivation} to three dimensions and a less trivial perturbation spectrum is straightforward. This implies that nonlinear phase-space echoes might be common features of galaxies.

On the other hand, just as in plasma physics, echoes will tend to be suppressed by small-scale diffusion, which in our case usually means orbital scattering by molecular clouds. This is because the echo relies on nonlinear coupling between wound-up phase space substructures. Therefore, there will be cases in which a very high resolution, `collisionless' $N$-body simulation of a galaxy will exhibit significant echoes, while the real `collisional' galaxy that is being modeled---or even a less well-resolved copy of the same $N$-body simulation---will not. What is more, this can be true even if the level of small-scale diffusion is nominally very weak. This is well-illustrated by Figs.\,\ref{fig:t2t1_dV2_Aecho_NoSSDiff} and~\ref{fig:t2t1_dV2_Aecho}, in which the echo signal is qualitatively changed, even though the level of small-scale diffusion would not heat the disc significantly on the timescale of interest.

We now discuss briefly the implications of our results for interpreting the vertical \textit{Gaia} Snail itself (\S\ref{sec:implication_GaiaSnail}). We then compare our work to some previous literature on nonlinear phase-space dynamics in galaxies (\S\ref{sec:Literature}).

\subsection{Implications for the \textit{Gaia} Snail}
\label{sec:implication_GaiaSnail}

Although our model is oversimplified in many respects, our simulation shows that two Sagittarius-like kicks can, in principle, generate an echo with amplitudes comparable to those of the \textit{Gaia} Snails ($A=0.1$--$0.2$; see \citealt{frankel_vertical_2023}). However, attributing the Snails we observe today entirely to echoes imposes rather stringent requirements. 

Suppose, quite reasonably, that the first kick was provided by the Sagittarius dwarf galaxy, which is expected to primarily excite an $n_1=1$ mode \cite[e.g.,][]{banik_comprehensive_2023}. Fig.\,\ref{fig:t2t1_dV2_Aecho} then implies that, for the echo to account for the \textit{Gaia} Snail, the second satellite/subhalo must have passed near the Solar neighborhood at a glancing angle $\sim 0.4 \Gyr$ after the passage of Sagittarius and delivered an antisymmetric kick of strength $\Delta V_2 \gtrsim 6 \kpc \Gyr^{-1}$, which requires a mass of $M_{\rm p} \gtrsim 10^8 \Msun$ (\S \ref{sec:model}, see also Appendix~\ref{sec:impulsive_satellite_perturbation} for the effect of time dependence, which further increases the required mass). However, there are no known luminous satellites of such mass near the Galactic disc apart from Sagittarius (see \citealt{banik_comprehensive_2022} and references therein). It is possible that the second kick was caused by an unseen dark subhalo, although such a massive subhalo would likely carry a detectable number of baryons. Moreover, several $N$-body simulations have shown that it is difficult to reproduce the Snail even by Sagittarius alone unless its {final mass is} much higher than is observed \citep[e.g.,][]{bland2021galactic,bennett2022exploring,Asano2025Ripples}. Since the (second-order) echo is inherently weaker than the (first-order) response probed in those studies, we expect that echoes would face even greater difficulty in reproducing the observed amplitude. 

A more realistic scenario would be that weak echoes from perturbations by Sagittarius and other satellites/subhalos are continually overlapping with the direct first-order responses, forming phase spirals with complicated variation in amplitude and winding time across both the vertical and horizontal phase space. The residuals between the data and models based on a single encounter indeed show coherent structures, indicating the presence of multiple overlapping phase-spirals \citep{frankel_vertical_2023}. Disentangling the precise origin(s) of the Snails remains a challenging task and lies far beyond the scope of this paper. On this front, various stellar labels like age and chemical abundances, which also exhibit Snail-like patterns, may offer valuable clues, as different labels have different gradients in vertical action and angular momentum, and can thus help constrain the nature of the perturbations \citep{Frankel2024Iron}.

\subsection{Relation to previous literature}
\label{sec:Literature}

It is instructive to contrast our work with a different nonlinear study of Snail formation by \cite{chiba2025origin}. They showed that the nonlinear response of a disc to \textit{persistent} resonant driving by spiral arms could produce asymmetric phase-space structures. In particular, they were able to produce a steadily-rotating, two-armed Snail in the case where they included small-scale orbital diffusion due to molecular cloud scattering. In other words, when coupled with a persistent nonlinear driving, small-scale diffusion can \textit{preserve}, rather than destroy, Snail-like phase space features. By contrast, in our study---as well as the stochastic model of \cite{tremaine_origin_2023}---the large-scale driving perturbations are very short lived. As a result, phase-space structures are always transient, and are always suppressed by small-scale orbital diffusion.

Previous papers displaying the results of $N$-body disc simulations have revealed phenomena resembling a `galactic echo' effect; i.e., a galaxy experiencing an internal perturbation sometime after experiencing two external perturbations \citep{chiueh_rotating_2000, fuchs_density_2005}. However, these papers do not explicitly draw a comparison to the plasma echo, nor do they work out a quantitative theory of when and how such an effect should arise.

Finally we note that our nonlinear echo effect is distinct from the mode-coupling investigated by \cite{tagger1987nonlinear}, \cite{sygnet1988non}, and \cite{masset1997non}. Those studies considered two (or more) concurrent spiral/bar perturbations in a disc, and the self-gravity of one such perturbation affected the evolution of the other, and vice versa. Our galactic echo, on the other hand, does not rely on self-gravity, but rather on the nonlinear coupling of phase-mixing features in the DF following two short-lived, independent, potential perturbations.

%%%%%%%%%%%%%%%%%%%%%%%%%%%%%%%%%%%%%%%%%%%%%%%%%%%%%%%%%%%%%%%%%%%%%%%%%%%%%%%%%%%%%%%%%%%%%%%%%%%%%%%%%%%%%%%%%%%%%%%%%%%%%%%%%%%%%%%%%%%%%%%%%%%%%%%%%%%%%%%%%%%%%%%%%%%%%%%%%%%%%%%%%%%%%%%%%%%%%%%%

\section{Summary}
\label{sec:Summary}

In this paper we have derived the galactic analogue of the plasma echo: a near-collisionless system (plasma or galaxy) is perturbed twice, and yet exhibits three macroscopic responses, with the third `echo' response arising due to a delayed nonlinear coupling between the first two. Our conclusions can be summarized as follows.

\begin{itemize}
    \item We found that the galactic version of the echo is very similar to the plasma one. The primary distinction lies in the fact that the potential fluctuations in a galaxy, when expressed in angle-action space, are functions of both canonical coordinates and momenta. Nevertheless, this difference does not introduce any qualitatively new phenomena. 
    \item We applied our echo theory to the vertical motion in the Milky Way and to the formation of the \textit{Gaia} Snails in vertical phase space. We found that vertical echoes are common and can easily dominate over the original responses at late times, since the latter are more tightly wound and hence more vulnerable to destruction by small-scale diffusion due to, e.g., molecular cloud scattering. However, the large amplitude of the observed Snail suggests that it is unlikely to be a pure echo effect. 
    \item The basic physical mechanism behind echoes is sufficiently generic and robust that we expect them to be common features in the (generally six-dimensional) phase-space of collisionless galactic discs. While the self-gravity of the perturbation is not essential for the formation of echoes, how it might affect the echo dynamics is unclear and remains a topic for future study.
\end{itemize}

%%%%%%%%%%%%%%%%%%%%%%%%%%%%%%%%%%%%%%%%%%%%%%%%%%%%%%%%%%%%%%%%%%%%%%%%%%%%%%%%%%%%%%%%%%%%%%%%%%%%%%%%%%%%%%%%%%%%%%%%%%%%%%%%%%%%%%%%%%%%%%%%%%%%%%%%%%%%%%%%%%%%%%%%%%%%%%%%%%%%%%%%%%%%%%%%%%%%%%%%

\section*{Data Availability}

%%%%%%%%%%%%%%%%%%%%%%%%%%%%%%%%%%%%%%%%%%%%%%%%%%%%%%%%%%%%%%%%%%%%%%%%%%%%%%%%%%%%%%%%%%%%%%%%%%%%%%%%%%%%%%%%%%%%%%%%%%%%%%%%%%%%%%%%%%%%%%%%%%%%%%%%%%%%%%%%%%%%%%%%%%%%%%%%%%%%%%%%%%%%%%%%%%%%%%%%

The numerical simulation results used in this paper will be shared on request to the corresponding author.

%%%%%%%%%%%%%%%%%%%%%%%%%%%%%%%%%%%%%%%%%%%%%%%%%%%%%%%%%%%%%%%%%%%%%%%%%%%%%%%%%%%%%%%%%%%%%%%%%%%%%%%%%%%%%%%%%%%%%%%%%%%%%%%%%%%%%%%%%%%%%%%%%%%%%%%%%%%%%%%%%%%%%%%%%%%%%%%%%%%%%%%%%%%%%%%%%%%%%%%%

\section*{Acknowledgements}

%%%%%%%%%%%%%%%%%%%%%%%%%%%%%%%%%%%%%%%%%%%%%%%%%%%%%%%%%%%%%%%%%%%%%%%%%%%%%%%%%%%%%%%%%%%%%%%%%%%%
We are grateful to N.~Bahcall, U.~Banik, J.~Goodman and M.~Medvedev for helpful feedback at various stages of this project. RC is supported by the Natural Sciences and Engineering Research Council of Canada (NSERC) (funding reference \#DIS-2022-568580) and the Japan Society for the Promotion of Science (JSPS) Research Fellowship, grant No. 25KJ0049. CH is supported by the John N.~Bahcall Fellowship Fund and the Sivian Fund at the Institute for Advanced Study. MWK was supported in part by NSF CAREER Award No. 1944972.

The authors thank the Kavli Institute for Theoretical Physics (KITP) for its hospitality during the completion of this work; KITP is supported in part by the National Science Foundation under Award No.~PHY-2309135. 

Computations were performed on the Niagara supercomputer at the SciNet HPC Consortium. SciNet is funded by Innovation, Science and Economic Development Canada; the Digital Research Alliance of Canada; the Ontario Research Fund: Research Excellence; and the University of Toronto.

\bibliographystyle{mnras}
\bibliography{bibliography}

%%%%%%%%%%%%%%%%%%%%%%%%%%%%%%%%%%%%%%%%%%%%%%%%%%%%%%%%%%%%%%%%%%%%%%%%%%%%%%%%%%%%%%%%%%%%%%%%%%%%%%%%%%%%%%%%%%%%%%%%%%%%%%%%%%%%%%%%%%%%%%%%%%%%%%%%%%%%%%%%%%%%%%%%%%%%%%%%%%%%%%%%%%%%%%%%%%%%%%%%
%% This command is needed to show the entire author+affiliation list when
%% the collaboration and author truncation commands are used.  It has to
%% go at the end of the manuscript.
%\allauthors

%% Include this line if you are using the \added, \replaced, \deleted
%% commands to see a summary list of all changes at the end of the article.
%\listofchanges

\appendix

%%%%%%%%%%%%%%%%%%%%%%%%%%%%%%%%%%%%%%%%%%%%%%%%%%%%%%%%%%%%%%%%%%%%%%%%%%%%%%%%%%%%%%%%%%%%%%%%%%%%%%%%%%%%%%%%%%%%%%%%%%%%%%%%%%%%%%%%%%%%%%%%%%%%%%%%%%%%%%%%%%%%%%%%%%%%%%%%%%%%%%%%%%%%%%%%%%%%%%%%

\section{Net kick by passing satellite in the impulsive regime}
\label{sec:impulsive_satellite_perturbation}

\begin{figure}
  \centering
  \includegraphics[width=1\linewidth]{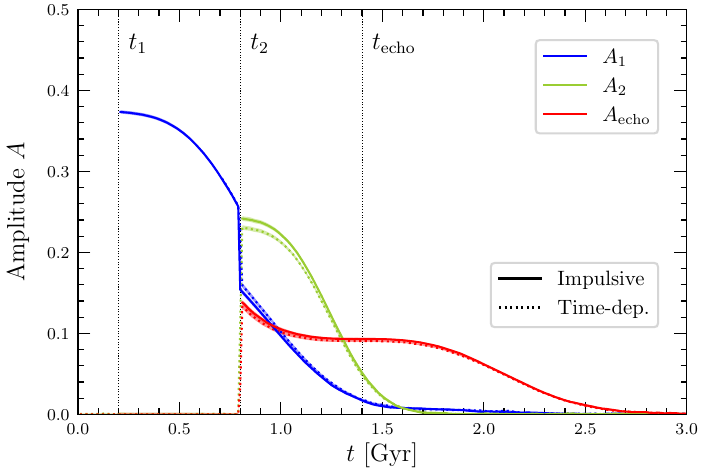}
  \includegraphics[width=1\linewidth]{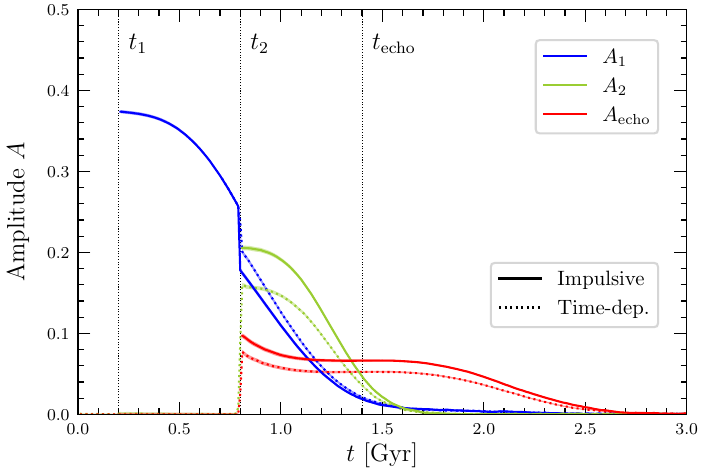}
  \caption{Amplitude of phase spirals as a function of time. The second perturbation at $t=t_2$ is modeled either as impulsive (solid) or time-dependent (dotted). The upper panel shows the fiducial case with $a=0.3\kpc$ and $M_{\rm p} = 1.2 \times 10^8 \Msun$, while the lower panel adopts a more extended and massive perturber with $a=1.5\kpc$ and $M_{\rm p} = 10^9 \Msun$. For larger $a$, the perturbation becomes less impulsive and hence the time-dependent effect becomes more pronounced.}
  \label{fig:t_A_timedep}
\end{figure}

In this appendix, we present our model for the net vertical kick induced by the second perturber on a low-inclination orbit using the impulse approximation and discuss its validity. For an earlier study on satellite perturbations in a 1D stellar disc model, see \cite{Weinberg1991Vertical}.

We consider the perturbation by a satellite galaxy and/or a dark-matter subhalo with mass $M_{\rm p}$ passing through the Solar neighborhood at time $t=t_{\rm p}$. We assume that the perturber is on a straight orbit with constant velocity $v_{\rm p}$ and incident angle $\vartheta_{\rm p}$ between the velocity vector and the $z$-axis. We further assume that the perturber is a Plummer sphere with core radius $a$. The potential of the perturber at the position of stars with height $z$ in the Solar neighborhood at time $\tau=t-t_{\rm p}$ from the impact time is
\begin{align}
  \Phi_{\rm p}(z,\tau) = - \frac{\rmG M_{\rm p}}{\sqrt{(v_{\rm p} \tau \sin \vartheta_{\rm p})^2 + (z - v_{\rm p} \tau \cos \vartheta_{\rm p})^2 + a^2}}.
  \label{eq:plummer_sphere_potential}
\end{align}
In the impulsive limit, we may assume that the stars move little while the perturber imparts a velocity kick. The net kick to stars in this limit is
\begin{align}
  &\Delta v(z) = - \int_{-\infty}^{\infty} \drm \tau \frac{\drm \Phi_{\rm p}}{\drm z} \nn \\
  &= - \int_{-\infty}^{\infty} \drm \tau \frac{\rmG M_{\rm p}(z - v_{\rm p} \tau \cos \vartheta_{\rm p})}{[(v_{\rm p} \tau \sin \vartheta_{\rm p})^2 + (z - v_{\rm p} \tau \cos \vartheta_{\rm p})^2 + a^2]^{3/2}}.
  \label{eq:plummer_sphere_kick_impulse}
\end{align}
When the perturber passes at a glancing angle $\vartheta_{\rm p} \approx \pi/2$, the kick reduces to
\begin{align}
  \Delta v(z) = - \int_{-\infty}^{\infty} \drm \tau \frac{\rmG M_{\rm p}z}{[(v_{\rm p} \tau)^2 + z^2 + a^2]^{3/2}} = - \frac{\rmG M_{\rm p}}{v_{\rm p}} \frac{2 z}{z^2 + a^2},
  \label{eq:plummer_sphere_kick_impulse_theta90}
\end{align}
which is an odd function of $z$, thus producing a quadrupole perturbation in phase space. The kick peaks at $z=\pm a$ with amplitude $\Delta V_2 = \rmG M_{\rm p}/(a v_{\rm p})$.

To test the validity of the impulse approximation, we compare simulations based on equation \eqref{eq:plummer_sphere_kick_impulse_theta90} with those that evolve under the fully time-dependent perturbation \eqref{eq:plummer_sphere_potential}. Throughout, we keep the first perturber impulsive and only investigate the time-dependent effect of the second. The upper panel of Fig.\,\ref{fig:t_A_timedep} shows the amplitude of the phase spirals for impulsive (solid) and time-dependent (dotted) perturbations using the default parameters (\S \ref{sec:model}): for the first perturber, $\Delta V_1 = 6 \kpcGyr$, and for the second, $a=0.3\kpc$, $M_{\rm p} = 1.2 \times 10^8 \Msun$, $\vartheta_{\rm p}=\pi/2$, and $v_{\rm p}=300 \kpcGyr$, which gives $\Delta V_2 \simeq 6 \kpcGyr$. The time-dependent perturbation produces nearly the same result as the impulsive case since the characteristic time of the perturbation is $a/v_{\rm p} = 1 \Myr$, which is much smaller than the stars' vertical oscillation period, $T \simeq 70 \Myr$, justifying the impulse approximation. The amplitude $A_2$ is slightly reduced, however, because stars rotate in phase space by a few times $(a/v_{\rm p})(2\pi/T) \sim 5^\circ$ during the passage of the perturber, so the perturbing force partially averages out. 

In our fiducial model, we assumed a rather compact second perturber ($a=0.3\kpc$, $M_{\rm p} = 1.2 \times 10^8 \Msun$) intended to represent a dark subhalo passing in the vicinity of the Solar neighborhood. For a Sagittarius-like perturber, both its size and mass may be larger. The lower panel of Fig.\,\ref{fig:t_A_timedep} shows results with $a=1.5\kpc$ and $M_{\rm p} = 10^9 \Msun$, with other parameters unchanged.\footnote{For reference, the present-day Sagittarius dwarf galaxy has mass $M_{\rm p} = 4 \times 10^8 \Msun$ and core radius of roughly $4^\circ$ on the sky, or $a \simeq 1.7 \kpc$ assuming a distance of $25\kpc$ \citep{Vasiliev2020Sagittarius}.} Despite the higher mass, the resulting perturbation $A_2$ is weaker than in the fiducial case (upper panel), since the kick is now strongest at $z = \pm 1.5 \kpc$, well beyond the disc's scale height ($h=0.2 \kpc$). The time-dependent perturbation still produces a sharp rise in amplitude, as $a/v_{\rm p} = 5 \Myr$ remains short compared to $T$, but it is suppressed significantly relative to the impulsive case because stars now advance in phase by a few times $(a/v_{\rm p})(2\pi/T) \sim 26^\circ$ while the perturber passes through.

\section{Nonlinear saturation of echo}
\label{sec:nonlinear_saturation_echo}

This appendix explores the behavior of the echo in the fully nonlinear (i.e., non-perturbative) regime. As in \S\ref{sec:echo} we ignore both the self-gravity of the fluctuations and any small-scale diffusion. Furthermore, we consider a homogeneous system, i.e., a system with no steady-state gravitational potential (so this calculation is very similar to the corresponding plasma calculation in \S IV of \citealt{ONeil1968Temporal}). The advantage of the homogeneous assumption is that we do not have to use angle-action variables.

Let us therefore consider a one-dimensional phase space $(x,v)$. The unperturbed distribution function is $f_0(v)$. Now we subject the system to a hammer that imparts a velocity change $\Delta v = k_1 \psi_1 \sin k_1 x$, arising from the potential $\Phi(x, t) = \delta(t - t_1) \psi_1 \cos k_1 x$ with $k_1 > 0$. The DF immediately after the kick is
\begin{align}
  f_\ast(x, v, t_1) = f_0(v - \Delta v) = f_0(v - k_1 \psi_1 \sin k_1 x).
\end{align}
At later times the DF is
\begin{align}
  f_\ast(x, v, t) = f_0\{v - k_1 \psi_1 \sin k_1[x - v(t - t_1)]\}.
\end{align}
At time $t_2$ we impose another kick $\Delta v = k_2 \psi_2 \sin k_2 x$ with $k_2 > 0$. The DF immediately after the second kick is
\begin{align}
  f(x, v, t_2) 
  &= f_\ast(x, v - \Delta v, t_2) \nn \\
  &= f_0\{v - k_2 \psi_2 \sin k_2 x \\
  &\hspace{9.6mm}- k_1 \psi_1 \sin k_1[x - (v - k_2 \psi_2 \sin k_2 x)(t_2 - t_1)]\} \nn
\end{align}
and at later times
\begin{align}
  f(x, v, t) = f_0\{v &- k_2 \psi_2 \sin k_2[x - v(t - t_2)] \nn \\
  &- k_1 \psi_1 \sin k_1[x - v(t - t_1) \nn \\
  &+ k_2 \psi_2 \sin\{k_2[x - v(t - t_2)]\}(t_2 - t_1)]\}.
  \label{eq:echo_solution_nonlinear}
\end{align}

We may expand equation (\ref{eq:echo_solution_nonlinear}) in powers of $\psi_1$ and $\psi_2$. Among the terms proportional to $\psi_1 \psi_2$, those that satisfy the echo condition ($t_{\rm echo} > t_2$) are
\begin{align}
  &f^{\,\rm echo}(x,v,t) = \frac{k_1 k_2 \psi_1 \psi_2}{2} 
  \bigg\{ 
    \frac{\pd^2 f_0}{\pd v^2} \cos\big[(k_2 - k_1)[x - v(t - t_{\rm echo})]\big] \nn \\
    & \hspace{10mm} - k_1(t_2 - t_1) \frac{\pd f_0}{\pd v} \sin\big[(k_2 - k_1)[x - v(t - t_{\rm echo})] \big]
  \bigg\},
  \label{eq:echo_solution_A1A2}
\end{align}
where $t_{\rm echo}=(k_2t_2-k_1t_1)/(k_2-k_1)$. As a consistency check, we note that this is equivalent to the second-order perturbative solution (\ref{eq:f_echo}) in a homogeneous system, i.e., when $\pd_J \psi_i = 0$ and $\pd_J \Omega$ is a constant.

We now compute the density at $x=0$ using the fully nonlinear solution (\ref{eq:echo_solution_nonlinear}):
\begin{align}
  \rho(0,t) = \int_{-\infty}^{\infty} \drm v f(0, v, t).
  \label{eq:rho_0}
\end{align}
We assume a Maxwellian for $f_0$, i.e., $(2\pi\sigma^2)^{-1/2} \exp [-v^2/(2\sigma^2)]$, in which case the density \eqref{eq:rho_0} could be expressed as a series expansion in terms of Bessel functions (\citealt{ONeil1968Temporal}). Here, we simply compute the velocity integral numerically.

\begin{figure}
  \centering
  \includegraphics[width=1\linewidth]{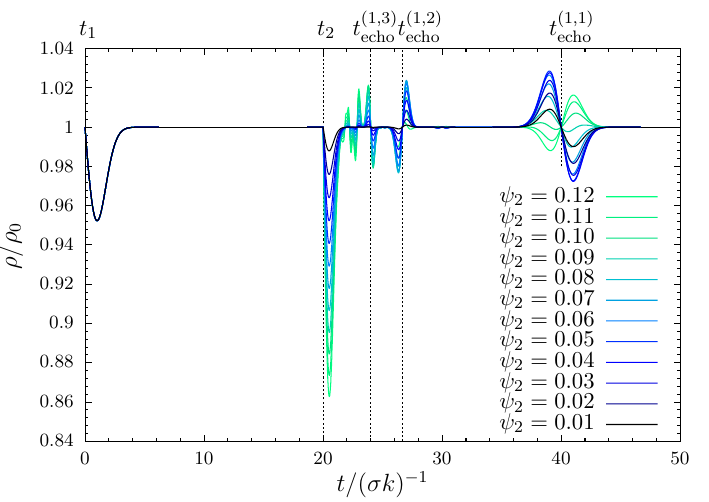}
  \caption{Density fluctuation at $x=0$ caused by two impulsive kicks at $t_1=0$ and $t_2=20(\sigma k)^{-1}$. The values of $\psi_2$ are given in units of $\sigma/k$.}
  \label{fig:rho_t0t20}
\end{figure}

Fig.\,\ref{fig:rho_t0t20} shows the density fluctuation \eqref{fig:rho_t0t20} caused by two impulsive perturbations with wave numbers $k_1=k$ and $k_2=2k$ at $t_1=0$ and $t_2=20(\sigma k)^{-1}$, where $\sigma$ is the velocity dispersion and $k=k_2-k_1$ is an arbitrary wave number; the second-order approximation \eqref{eq:echo_solution_A1A2} would then predict an echo at $t_\mathrm{echo} = 40 (\sigma k)^{-1}$. We fixed the amplitude of the first impulse at $\psi_1=0.1 \sigma/k$ and varied that of the second impulse, $\psi_2$, indicated by the different colors.

Let us first focus on the black line, which is for a weak second kick, $\psi_2 = 0.01 \sigma/k$. As in Fig.\,\ref{fig:meanz}, the density undergoes brief fluctuations immediately following the two kicks, which rapidly damp out as the system phase mixes. Then, near $t = 40 (\sigma k)^{-1}$, the phase-space perturbation unwinds and thereby generates a renewed density fluctuation, as expected from the second-order solution \eqref{eq:echo_solution_A1A2}. However, as we increase $\psi_2$, we observe two phenomena that are not captured by the second-order perturbation theory. 

First, the density shows multiple echo-like structures just after the second impulse. Since these features are absent at small $\psi_2$, they are most likely higher-order echoes. While one could in principle derive explicit solutions for these echoes by expanding \eqref{eq:echo_solution_nonlinear} in powers of $\psi_1$ and $\psi_2$, this becomes increasingly tedious at higher orders. However, the \textit{timing} of these higher-order echoes can be inferred as follows. Each power of $\psi_i$ is associated with sines and cosines with argument $\phi_i \equiv k_i [x - v(t - t_i)]$. The terms of order $l$ have a factor $\psi_1^m \psi_2^{l-m}$ where $m = 1,...,l-1$ (the terms with $m = 0$ and $m = l$ arise from a single hammer and therefore do not produce an echo). These terms involve $\sin^m \phi_1 \sin^{l-m} \phi_2$ and analogous cosine terms and so produce terms like $\sin (j_1 \phi_1) \sin (j_2 \phi_2)$ where $0 \leq j_1 \leq m, 0 \leq j_2 \leq l - m$ and analogous cosine terms. Expanding these products of sines and cosines, we get oscillatory terms with argument $j_1 \phi_1 \pm j_2 \phi_2$. These can produce echoes at $t =  t^{(j_1,j_2)}_\mathrm{echo}$ where
\begin{align}
  t^{(j_1,j_2)}_\mathrm{echo} = \frac{j_2 k_2 t_2 \pm j_1 k_1 t_1}{j_2 k_2 \pm j_1 k_1} = t_2 \mp \frac{j_1 k_1 (t_2 - t_1)}{j_2 k_2 \pm j_1 k_1}.
\end{align}
With the upper sign, the second term is always negative, so the echo appears before $t_2$ and thus is unphysical [c.f. equation \eqref{eq:echo_time_unphysical}]. Therefore echoes occur only at
\begin{align}
  t^{(j_1,j_2)}_\mathrm{echo} = \frac{j_2 k_2 t_2 - j_1 k_1 t_1}{j_2 k_2 - j_1 k_1} = t_2 + \frac{j_1 k_1 (t_2 - t_1)}{j_2 k_2- j_1 k_1},
  \label{eq:t_high_order_echo}
\end{align}
with constraints $k_1, k_2 > 0$, $j_1, j_2 > 0$, $j_1 + j_2 \leq l$ and $j_2 k_2 > j_1 k_1$. As a verification, we marked in Fig.\,\ref{fig:rho_t0t20} the timing of the standard second-order echo $(j_1,j_2)=(1,1)$, the third-order echo $(j_1,j_2)=(1,2)$, and the fourth-order echo $(j_1,j_2)=(1,3)$. Note that contrary to other nonlinear effects, which typically become prominent at late times, the echo appears earlier the more nonlinear it is (higher $l$).

Another nonlinear feature found at large $\psi_2$ is the saturation of the second-order echo. Precisely, as $\psi_2$ is increased, the echo around $t = 40(\sigma k)^{-1}$ initially grows in amplitude but then saturates and decays---a behavior also observed in our (inhomogeneous) simulations (Fig.\,\ref{fig:t2t1_dV2_Aecho_NoSSDiff}). Interestingly, at even larger $\psi_2$, the echo grows again but the density perturbation flips sign. This behavior results from the higher-order terms with $j_1=j_2$, which produce echoes with the same timing as the second-order echo. At each order $l$, the dominant contribution comes from terms proportional to $\psi_1 \psi_2^{l-1}$, since these include a term proportional to $(t_2-t_1)^{l-1}$ and $t_2-t_1$ is a large parameter. Taking the $l$th derivative of (\ref{eq:echo_solution_nonlinear})---once with respect to $\psi_1$ and $l{-}1$ times with respect to $\psi_2$--and retaining only terms proportional to $(t_2-t_1)^{l-1}$, we find that
\begin{align}
  &f_l(x,v,t) 
  = - \frac{k_1 \psi_1 (k_2 \psi_2)^{l-1}}{(l-1)!} \frac{\pd f_0}{\pd v} (\tau \sin \phi_2)^{l-1} \frac{\pd^{l-1}}{\pd \phi_1^{l-1}}\sin \phi_1,
  \label{eq:f_high_order}
\end{align}
where $\phi_i \equiv k_i [x - v(t - t_i)]$ and $\tau \equiv k_1 (t_2-t_1)$. Among the various terms that arise from the power reduction of $(\sin \phi_2)^{l-1}$, those that generate echoes at the 2nd-order echo time ($j_1=j_2$) are
\begin{align}
  f_l(x,v,t) = (- 1)^{l/2} \frac{k_1 \psi_1 (k_2 \psi_2 \tau / 2)^{l-1}}{(l-1)!} \binom{l-2}{\frac{l-2}{2}} \frac{\pd f_0}{\pd v} \sin(\phi_2-\phi_1),
  \label{eq:f_high_order_j1}
\end{align}
where $l$ is even. The 2nd-order term ($l=2$) yields the second term of (\ref{eq:echo_solution_A1A2}). Due to the alternating sign $(- 1)^{l/2}$ in (\ref{eq:f_high_order_j1}) the 4th-order term ($l=4$) weakens the 2nd-order echo, causing it to saturate and decay at large $\psi_2 \tau$. The amplitude of the sum of the two flips sign when
\begin{align}
  - \frac{(k_2 \psi_2 \tau / 2)}{1!} \binom{0}{0} + \frac{(k_2 \psi_2 \tau / 2)^3}{3!} \binom{2}{1} > 0 ~~~~
  \therefore \psi_2 > \frac{\sqrt{12}}{k_2\tau} \sim 0.09,
\end{align}
roughly consistent with the numerical result (Fig.\,\ref{fig:rho_t0t20}). Equation (\ref{eq:f_high_order_j1}) implies that the density perturbation caused by the echo will continue to deform with increasing values of $\psi_2 \tau$.

\bsp
\label{lastpage}
\end{document}